# Fabrication framework for three-dimensional colloidal particles with decoupled geometry and material composition

Hamed Almohammadi[a*], Gurminder K. Paink[a], Vasumathi Venkat[a,b], Jacopo Movilli[a,c], Atalaya Milan Wilborn[a], Haritosh Patel[a], Haichao Wu[a], David Weitz[a,d], Joanna Aizenberg[a,e]

a. John A. Paulson School of Engineering and Applied Sciences, Harvard University, Cambridge, MA, USA
b. Department of Electrical and Computer Engineering, Princeton University, Princeton, NJ, USA
c. Department of Chemical Sciences, University of Padua, Padua, Padova, Italy
d. Department of Physics, Harvard University, Cambridge, MA, USA
e. Department of Chemistry and Chemical Biology, Harvard University, Cambridge, MA, USA

*Corresponding author: halmohammadi@seas.harvard.edu

**Abstract**

Shape-programmable particles offer significant opportunities for microrobotic systems at the individual level and for hierarchical materials with emergent functionalities arising from collective particle behavior. However, fabricating shape-changing stimuli-responsive particles with complex three-dimensional geometries at colloidal length scales remains a major challenge. Here, we introduce a general fabrication framework that decouples particle geometry from material composition to produce free-standing three-dimensional colloidal particles with complex architectures. Our approach combines soft lithography, swelling-assisted extraction, and sacrificial adhesive transfer to fabricate particles with high geometric fidelity. We establish a predictive framework that defines the accessible design space for increasingly complex particle geometries. We further extend our framework to fabricate high-aspect-ratio pillar arrays with intricate three-dimensional architectures at colloidal length scales. To demonstrate material versatility, we fabricate particles from both liquid crystal elastomers and hydrogels. We show reversible shape-changing behavior of liquid crystal elastomer particles with cylindrical and chiral shapes under thermal and optical stimuli. In suspension, these particles display collective optical dynamic behavior arising from coupling between changes in the programmed liquid crystal organization within the particles and stimulus-induced geometric reconfiguration of the particles. Collectively, this work establishes a versatile platform for geometry-programmable colloidal particles with emergent collective functionalities, providing a route toward materials and fluids with dynamically programmable properties.

## 1. Introduction

Programming particle geometry offers a powerful route to engineer matter across scales, enabling functionalities that emerge from individual particle behavior and collective assembly.[1-3] At the single-particle level, shape-programmable particles have enabled microscale robotic systems for applications such as targeted drug delivery, cargo transport, microsurgery, and biosensing.[4] Beyond individual functionality, particle shape also plays a central role in determining collective behavior. When these particles are dispersed in a fluid, their emergent collective functionalities offer promising directions for realizing fluids with programmable properties, including rheological and optical responses, also called programmable fluids or metafluids.[5,6] At the assembly level, particle architecture has emerged as a key design parameter for directing colloidal self-assembly and engineering complex assembly states[7-9], with reconfigurable particle geometries providing opportunities to access new assembly states that have been explored mainly in theoretical studies.[9–11] Despite these opportunities, fabricating shape-changing particles with arbitrary three-dimensional architectures at colloidal length scales remains a significant challenge. At these dimensions, thermal fluctuations dominate over gravitational forces, enabling particles to explore their accessible configuration space, self-organize, and dynamically reconfigure into collective states. These characteristics are essential for the development of programmable fluids and hierarchical functional materials.

Existing fabrication approaches for stimuli-responsive particles either constrain geometry or intrinsically couple geometric design to material composition. These methods are particle stretching and deformation[12–15], fiber-to-particle fabrication approaches[16-17], droplet-based microfluidic synthesis[18–22], soft lithography[23-24], and two-photon polymerization (2PP)[25–28]. While these approaches have enabled significant advancements in stimuli-responsive particles, the accessible particle geometries are often highly constrained by underlying fabrication processes, including deformation pathways, fiber architecture, flow-generated geometries, or planar templating, making it nearly impossible to fabricate particles with arbitrary three-dimensional geometries at colloidal length scales. Among existing fabrication approaches, 2PP overcomes these geometric constraints, enabling the fabrication of highly precise three-dimensional particles with submicron resolution.[25–28] However, this approach requires printing of the desired structure directly from a compatible printable material composition, intrinsically coupling geometric design to material selection. Consequently, it remains a major challenge to fabricate free-standing colloidal particles with arbitrary three-dimensional shapes while independently controlling geometry and material composition.

Here, we introduce a general fabrication framework that decouples particle geometry from material composition, enabling the fabrication of increasingly complex three-dimensional colloidal particles from stimuli-responsive materials. We further establish predictive design rules that define the accessible design space for increasingly complex particle geometries. We use this platform to fabricate liquid crystal elastomer (LCE) and hydrogel particles, as representative model systems of stimuli-responsive materials, to demonstrate the material compatibility of our approach. We demonstrate that LCE particles undergo reversible geometric changes in response to thermal and optical stimuli. These single-particle level transformations result in emergent collective optical behavior in particle ensembles through coupled changes in the programmed liquid crystal organization within the particles and their geometry. By enabling independent control over particle geometry and material composition, our fabrication approach provides a novel direction for stimuli-responsive colloidal matter, whose macroscopic collective functionalities emerge from particle-level shape transformations.

## 2. Results and Discussion

### 2.1 Fabrication of free-standing three-dimensional colloidal architectures

Our fabrication strategy (Figure 1a) combines the high geometric precision and design freedom of 2PP with a materials-independent soft lithography replication process, thereby decoupling particle geometry from material composition. In the soft lithography process, a reusable inverse PDMS (polydimethylsiloxane) mold replicated from a master template serves as a materials-independent template, into which the material of choice is cast, polymerized, and subsequently demolded to obtain the replicated structure. Applying this strategy to fabricate colloidal particles with increasingly complex three-dimensional shapes, however, presents two distinct fabrication challenges.

The first challenge arises during demolding of the replicated structures from the inverse PDMS mold. As particle geometry becomes increasingly complex, features such as overhangs and negative tapering mechanically interlock with the surrounding PDMS during demolding, hindering particle release and increasing the risk of structural damage. The second challenge is converting the replicated structures into free-standing particles. The replicated structures remain connected through a supporting substrate and therefore exist as substrate-attached pillar arrays rather than individual colloidal particles. Separating these structures into individual free-standing colloidal particles while preserving their three-dimensional geometry presents another fabrication challenge.

Our fabrication framework addresses these challenges independently. To overcome the first challenge, we employ swelling-assisted extraction by temporarily swelling the PDMS mold with a selective solvent, thereby increasing the clearance around geometrically interlocked features. In Section 2.2, we establish a predictive framework that defines geometry-dependent design principles for particle extractability. To overcome the second challenge, rather than releasing substrate-attached pillars, our strategy isolates the replicated structures within the PDMS mold and extracts them directly as free-standing colloidal particles using a sacrificial adhesive transfer layer.

Figure 1a summarizes the fabrication workflow, and Figure 1b shows a representative chiral epoxy pillar particle prepared by 2PP together with its corresponding replica made from LCE[29-31]. Epoxy master templates are fabricated on glass substrates using commercially available 2PP systems and subsequently replicated into an inverse PDMS mold (see also Methods and materials). Because the master templates are reusable, the fabrication strategy enables repeated material replication without repeated 2PP printing. PDMS is cast over the master templates and cured to form the inverse mold containing cavities corresponding to the intended particle geometry, which is subsequently swollen with hexane prior to demolding. Following solvent removal, the PDMS recovers its original dimensions and serves as a reusable mold. The desired precursor is introduced by capillary infiltration and polymerized according to the material-specific protocol. Excess precursor is removed by shear, leaving isolated structures embedded within the PDMS cavities. The mold is subsequently reswollen, after which a sacrificial water-soluble adhesive film is applied to extract the embedded particles directly from the cavities (Figure 1c). Dissolution of the adhesive film in water releases the particles into suspension (Figure 1d). Analysis of the particle trajectories reveals a mean-squared displacement (MSD) that scales linearly with lag time ($\Delta t$: time interval between particle positions), consistent with Brownian motion (Figure 1e, Supplementary Movie 1-2), confirming that the released structures behave as colloidal particles.

The same templated transfer strategy can also be used to fabricate high-aspect-ratio pillar arrays. In this case, the replicated pillars are demolded directly from the PDMS mold while preserving their intricate three-dimensional geometry. Prior to demolding, the PDMS mold is temporarily reswollen with hexane to reduce interfacial adhesion and minimize structural damage (see Methods and Materials and Fig. S1).

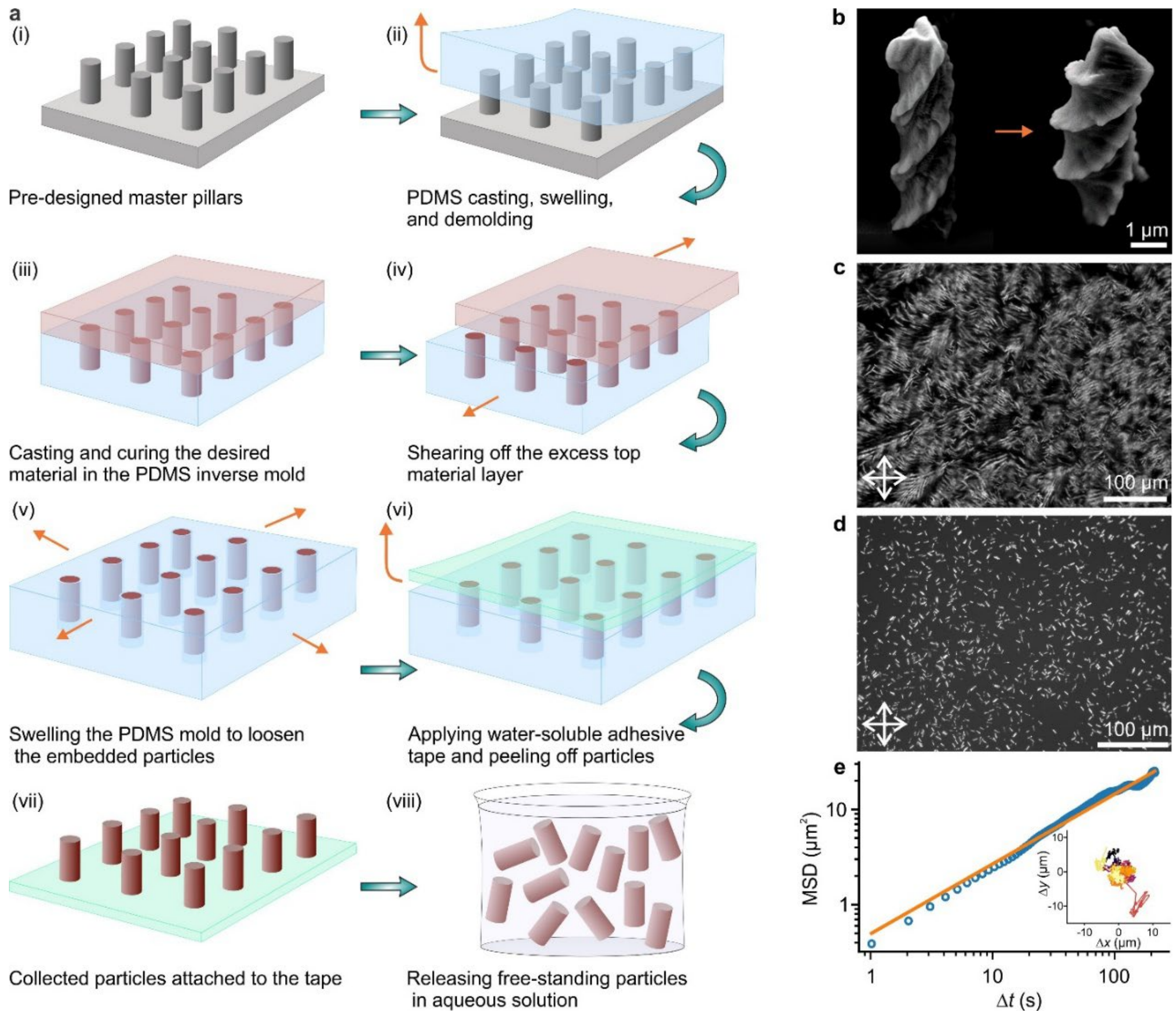

**Fig. 1. Decoupling geometry from the material composition in the fabrication of free-standing three-dimensional colloidal particles. a**, Schematic illustration of the fabrication workflow: (i) fabrication of predesigned master pillars by two-photon polymerization (2PP), (ii) polydimethylsiloxane (PDMS) casting and demolding following PDMS swelling to facilitate nondestructive release, (iii) casting and curing of the target material in the PDMS inverse mold, (iv) shearing off the excess top material layer to leave only particles embedded in the PDMS mold, (v) swelling the PDMS mold to loosen the embedded particles, (vi) applying water-soluble adhesive tape and peeling off the particles, (vii) collecting the particles on the tape, and (viii) releasing the free-standing particles into aqueous solution. **b**, Scanning electron microscope (SEM) image of a chiral pillar fabricated by 2PP (left) and its corresponding replica made from stimuli-responsive liquid crystal

elastomer (LCE) (right). **c**, Crossed-polarizer image of cylindrical LCE particles (length: 15 μm, diameter: 3 μm) attached to the adhesive tape before release, corresponding to step (vii) in panel (**a**). White arrows indicate the polarization directions. **d**, Crossed-polarizer image of cylindrical LCE particles (length: 10 μm, diameter: 2 μm) after release into water. White arrows indicate the polarization directions. **e**, Mean-square displacement of the particles (**d**) versus lag time, showing a linear dependence. The solid line is a linear fit ($R^2$=0.97). The inset shows the in-plane trajectories of the tracked particles.

### 2.2 Geometry-dependent design principles for particle extraction

We next develop a predictive extraction framework that relates particle geometry, particle-PDMS mold adhesion, and solvent-induced PDMS swelling to the mechanics of particle release (Figure 2). We classify the extraction mechanics into two regimes based on changes in the cross-section size of the particle geometries along the release direction in Figure 2a. Specifically, we consider geometries satisfying $\partial r/\partial z \geq 0$ and those for which $\partial r/\partial z < 0$. This distinction determines whether particle extraction is governed primarily by interfacial adhesion between the PDMS mold and the particle or by a combination of interfacial adhesion and geometric mechanical interlocking with the surrounding PDMS mold.

In the first regime ($\partial r/\partial z \geq 0$), the particle geometry lacks overhanging features in the extraction direction. Consequently, particle release is predominantly governed by the balance of adhesive interactions between the particle and the PDMS mold. Successful extraction requires that the applied force, $F_{\text{pull}}$, exceeds the particle-PDMS mold adhesion, $F^{\text{PDMS}}{}_{\text{adh}}$, while remaining below both the adhesion force between the particle and transfer substrate, $F^{\text{tape}}{}_{\text{adh}}$, and the mechanical failure threshold of the particle itself. Here, solvent-induced swelling of PDMS reduces particle–mold adhesion by expanding the PDMS network, thus creating an interfacial gap with a thickness of $h$ between the particle and the mold.

In the second regime ($\partial r/\partial z < 0$), particle geometries exhibit local increases in cross-sectional radius along the extraction direction, resulting in geometric overhangs and mechanical interlocking with the surrounding PDMS mold. Here, in addition to the particle-mold interfacial adhesion, $F^{\text{PDMS}}{}_{\text{adh}}$, the ability to elastically deform the mold around the particle is necessary for successful particle release. The relevant geometric parameter describing the degree of interlocking is defined as the difference between the maximum and minimum radial dimensions ($d = r_{\max} - r_{\min}$) of the particle at a given angle $\theta$, as illustrated schematically for chiral particles in Figure 2b. When $d/h \leq 1$, solvent-induced swelling of the mold generates sufficient free volume for the particle extraction to proceed similarly to regime one. In contrast, when $d/h > 1$, successful particle extraction requires elastic deformation of the surrounding PDMS mold. The value of $h$ depends on the choice of solvent. Here, we use hexane because nonpolar hydrocarbons strongly swell PDMS due to their similar solubility parameters[32]. To describe the swelling dynamics, we approximate the characteristic PDMS expansion as $l(t)/l_0 = 1 + S_\infty(1 - e^{-t/\tau})$, where $l(t)$ and $l_0$ are, respectively, PDMS mold dimension at time $t$ and its initial dimension, $S_\infty$ ($= (l_\infty - l_0)/l_0$) is the equilibrium swelling strain with $l_\infty$ being the PDMS mold dimension at the equilibrium, and $\tau$ is the characteristic solvent uptake timescale. At long times ($t \gg \tau$) the swelling approaches an equilibrium expansion ratio $l(t)/l_0 = 1 + S_\infty$, Figure 2c. For a particle of radius $r$, writing $(r+h)/r = 1 + S_\infty(1 - e^{-t/\tau})$, swelling generates an effective release gap of $h(t) = rS_\infty(1 - e^{-\frac{t}{\tau}})$, with an equilibrium value for $h$ as $h_\infty = rS_\infty$. For the PDMS formulation used in this work (Methods and materials), we measured $S_\infty$ to be 0.32, resulting in $h_\infty = 0.32r$, Figure 2c. A comprehensive comparison of PDMS swelling ratios in different solvents is provided in Ref. 32. This value of $h_\infty = 0.32r$ satisfies $d/h \leq 1$ when $d < 0.32r$, providing a predictive design criterion that determines whether swelling alone is sufficient for particle extraction (regime one). We apply this criterion to various particle geometries in Section 2.3.

Together, this framework establishes quantitative design rules linking particle geometry to extractability, providing a general guideline for fabricating increasingly complex three-dimensional colloidal particles.

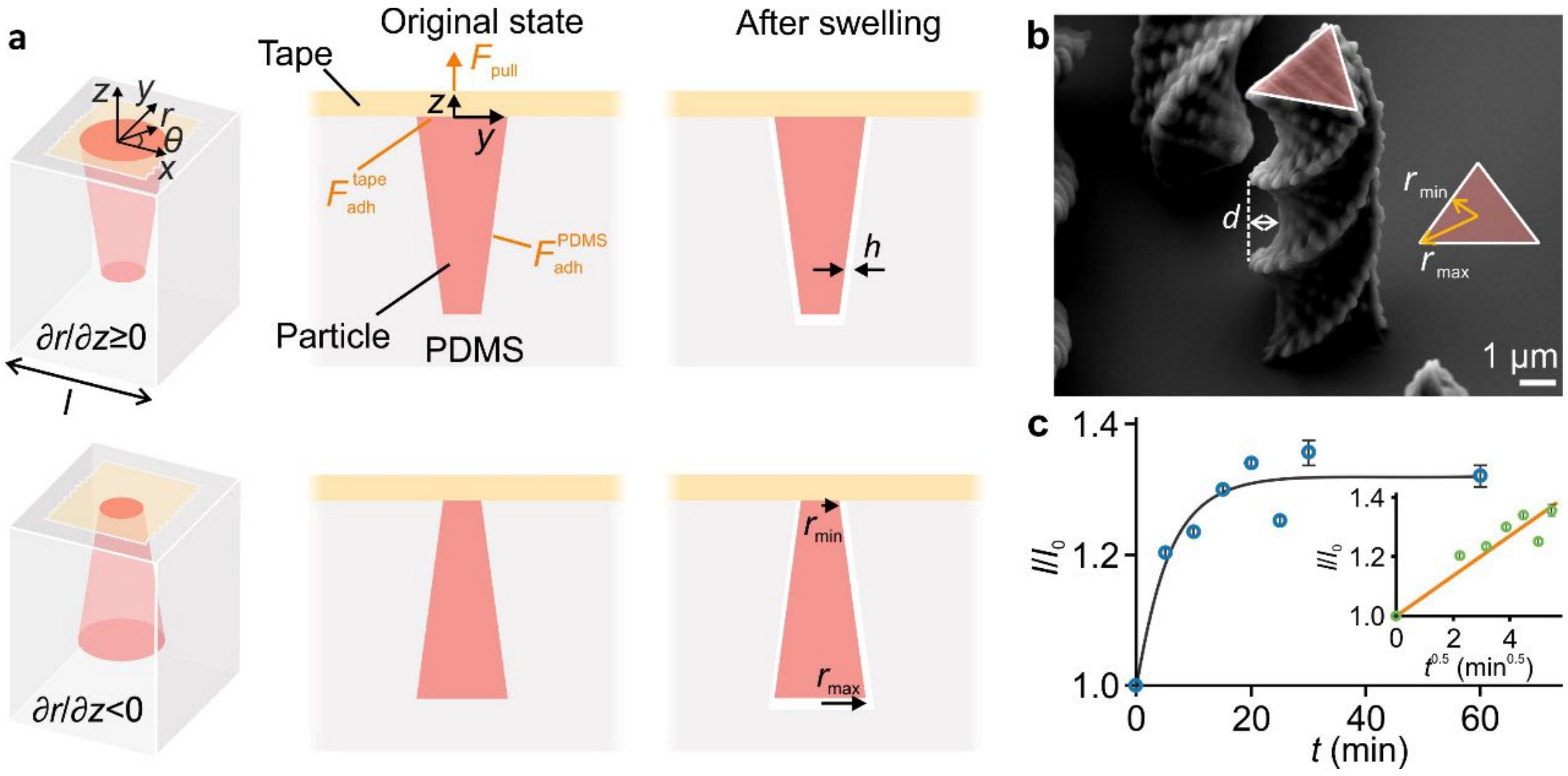

**Fig. 2. Design rules governing the extraction of free-standing particles from PDMS. a,** Schematic illustration of the mechanical and geometrical parameters governing particle extraction from the PDMS mold. The particle cross-sectional geometry is defined as $r(\theta, z)$. The extraction process is governed by the competition between the pulling force $F_{pull}$, the adhesion force between the tape and particle, $F^{tape}{}_{adh}$, and the adhesion force between the particle and PDMS, $F^{PDMS}{}_{adh}$. Swelling of the PDMS creates an interfacial gap. $h$, between the particle and the surrounding PDMS mold. **b,** SEM image of a chiral pillar highlighting the geometrical features relevant to particle extraction and release from the PDMS mold. **c,** Measured swelling-induced expansion of PDMS, $l/l_0$, as a function of immersion time, $t$, in hexane. Here, $l$ and $l_0$ are, respectively, the PDMS mold dimension at a time $t$ and the initial time. The inset shows the linear dependence of the $l/l_0$ on, $t^{1/2}$, consistent with diffusion-dominated solvent uptake. To quantify the swelling kinetics, the data are fit using $l(t)/l_0 = 1 + S_\infty(1 - e^{-t/\tau})$, with $S_\infty$ $(= (l_\infty - l_0)/l_0)$), the equilibrium swelling strain, $\tau$, the characteristic solvent uptake timescale, and $l_\infty$ the PDMS mold dimension at the equilibrium. The values for $\tau$ and $S_\infty$ are found to be 5.8 min and 0.32, respectively.

### 2.3 Experimental realization of complex particle geometries

We next experimentally test the geometry-dependent design space predicted by the extraction framework. Figure 3a presents a systematic series of three-dimensional LCE structures with progressively increasing geometric complexity, including positive-tapering, anisotropic, polygonal faceting, and negative-tapering particles with interlocking features (see also Fig. S2 for twisted chiral particles). For clarity, the structures shown in Figure 3a are imaged as substrate-attached pillars, which enable higher-resolution visualization of the complex three-dimensional geometries by scanning electron microscopy (SEM). In the following panels, we present the corresponding particles after their release into suspension.

Among the tested geometries, twisted chiral and negative-tapering particles represent the most demanding tests of the extraction framework due to geometric interlocking with the surrounding PDMS. Figure 3b shows a representative SEM image of a chiral pillar array and a bright field microscope image of the same particles suspended in silicone oil after release (Methods). The dimensions of released particles are measured to be 14.60 ± 0.90 μm in length and 3.21 ± 0.29 μm in diameter as opposed to the designed dimensions of 15 μm and 3 μm (Figure 3b). These data show deviations of only 2.7% and 7.1%, in length and diameter, respectively, which indicate a close match between the intended and fabricated dimensions, demonstrating that complex three-dimensional features can be faithfully preserved during extraction. Similarly, Figure 3c, shows a representative SEM and bright-field microscope images of a tapered pillar array and the corresponding released particles suspended in silicone oil, respectively. The lengths and tapering angles of the released particles are measured to be 10.19 ± 0.51 μm and 5.77 ± 1.18°. Compared to the designed dimension 10 μm length and tapering angle of 5.71°. These measurements correspond to deviations of only 1.9% in length and 1.1% in tapering angle. Together with the chiral particles, these measurements demonstrate that the framework preserves geometric features across distinct classes of particle architectures.

The predictive power of the extraction framework can be assessed by comparing its predicted extraction regimes with the experimentally fabricated particle geometries. The first five particle classes in Fig. 3a lie within the adhesion-dominated regime ($d<0.32r$), where the framework predicts that swelling-assisted disengagement alone should be sufficient for successful extraction. In contrast, the three negative-tapering geometries ($d=0.40r$, $d=1.00r$, and $d=1.33r$; top to bottom rows in Fig. 3a) were intentionally designed to exceed the criterion of $d=0.32r$. The chiral particles ($d=0.67r$; Fig. 3b) also fall within this deformation-assisted regime. Together, these geometries provide direct experimental tests of the deformation-assisted extraction regime, demonstrating that elastic deformation of the swollen PDMS substantially extends the range of extractable particle geometries. The largest tested geometry ($d=1.33r$) approaches the practical extraction limit of the present fabrication strategy, suggesting that the accessible design space is ultimately constrained by the finite deformability of the swollen PDMS. Taken together, these experiments demonstrate that the proposed framework outlines the extractability of complex particle geometries and provides practical guidelines for designing free-standing three-dimensional colloidal particles.

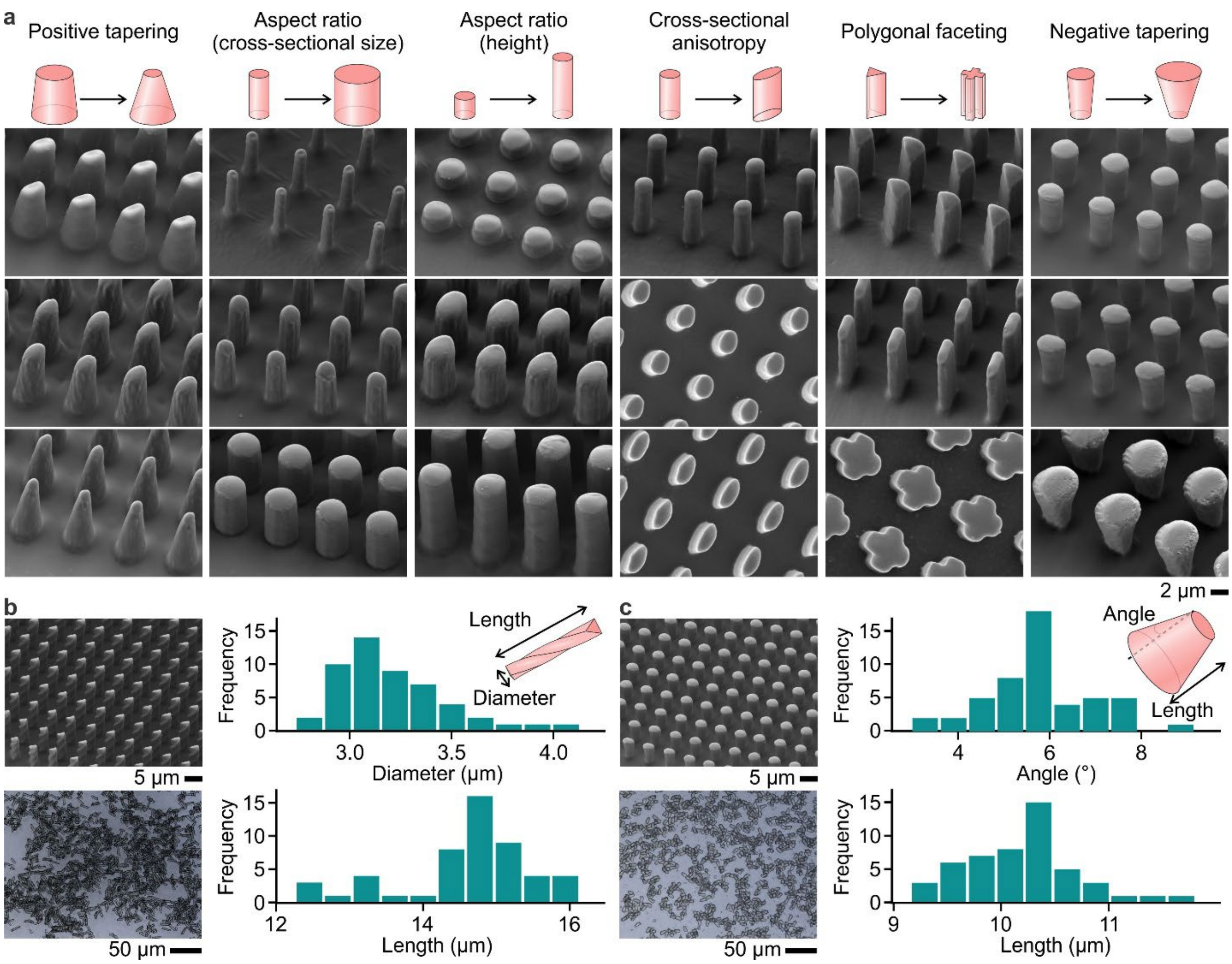

**Fig. 3. Experimental realization of complex particle geometries. a**, SEM images of three-dimensional LCE structures with geometries systematically spanning increasing geometric complexity, including independently tapered shapes, tunable cross-sectional size, longitudinal length, anisotropy, polygonal faceting, and negative-tapering architectures. Structures are shown in the pillar state prior to release for improved visualization, with selected geometries displayed from the top view to highlight cross-sectional features. **b**, SEM image (top left) of chiral particle array and bright-field microscope image (bottom left) of the particles suspended in silicone oil. The measured dimensions of the fabricated particles are 14.60 ± 0.90 µm in length and 3.21 ± 0.29 µm in diameter, compared with the intended dimensions of 15 µm and 3 µm, respectively. **c**, SEM image (top left) of tapered particle array and bright-field microscope image (bottom left) of the tapered particles suspended in silicone oil. The fabricated particles have a measured length of 10.19 ± 0.51 µm and tapering angles of 5.77 ± 1.18°, compared with the intended values 10 µm and 5.71°, respectively.

**2.4 Stimuli-responsive shape-shifting colloidal particles**

Having established the framework defining the accessible design space for increasingly complex particle geometries, we next demonstrate its material versatility by fabricating stimuli-responsive particles from LCEs and hydrogels (Figure 4). To fabricate LCE particles, the LCE precursor solution is cast into a PDMS mold with the geometry of interest and subsequently polymerized in the presence of an external magnetic field which aligns the liquid crystal molecules along the long axis of the particle, resulting in a well-defined director corresponding to the average molecular alignment direction (see Methods and materials for preparation details and Figure S3 for magnetic alignment)[33-34]. Once released, the particles are resuspended in silicone oil.

We first examine thermally induced shape transformations of the fabricated LCE particles. Figure 4a shows representative cylindrical and chiral particles before and after actuation (Supplementary Movie 3). Heating above the nematic-to-isotropic (ordered-to-disordered) transition temperature disrupts ordered molecules within the LCE, resulting in contraction along the director field of the molecules within the particles. These changes result in a decrease in the aspect ratio (*AR*) of both cylindrical and chiral particles. Figure 4b shows the normalized aspect ratio, $AR/AR_0$, for diameters ranging from 1 to 6 μm at a fixed particle length of 15 μm. Here, $AR_0$ stands for initial aspect ratio of particles. Across both particle geometries and various diameters, thermal actuation results in $AR/AR_0 \approx 0.6$. Next, we examine the photoresponse of the chiral LCEs particles. Under ultraviolet light exposure, trans–cis photoisomerization of the azobenzene moieties locally disrupts the ordering of the molecules within the particles, inducing reversible shape deformation (Figure 4c). These changes result in the decrease of the aspect ratio to $AR/AR_0=0.80\pm0.07$ for particles with a diameter of 3 μm and a length of 15 μm, Fig. 4d.

After confirming that the LCE particles individually exhibit stimuli responsive behavior, we next investigated the collective optical response of particle suspensions after thermal actuation. Figure 4e shows suspensions of chiral particles in silicone oil and imaged under crossed polarizers during thermal actuation (Supplementary Movie 4). Upon actuation, the suspension goes from a bright birefringent state to a darker state as the orientational order of liquid crystal molecules decreases, consistent with the transition from nematic to isotropic phase change. Quantitative image analysis reveals an approximately 24% reduction in optical intensity under crossed polarizers. In parallel, bright-field imaging shows an increase of approximately 5% in average image brightness (Figure 4f). This increase is due to the shape deformation, which results in a decrease in the projected particle area in the imaging plane as the particles predominantly lie parallel to the substrate. Our measurement for a chiral particle with an initial length of 15 μm and a diameter of 3 μm, $AR/AR_0=0.80\pm0.07$ upon actuation (Fig. 4d), indicates an approximately 14% increase in open projected area within the field of view when lying parallel to the substrate.

Finally, to demonstrate broader material compatibility beyond LCEs, we fabricated pHEMA-based hydrogel[35] colloidal particles, using the same strategy (Figure 4g-h). The fabricated cylindrical particles in a dehydrated state have a length of 9.92±0.46 μm in close agreement with the intended dimension of 10 μm (0.7% deviation), again demonstrating the dimensional accuracy of our framework. While the swelling behavior of the particles was not investigated here, pHEMA-based hydrogels are known to undergo reversible isotropic swelling in aqueous environments, with reported equilibrium volume swelling ratios in the range of approximately 1.5–1.8 depending on crosslink density and network composition[35]. Collectively, these results show that the same fabrication framework is applicable to different responsive materials.

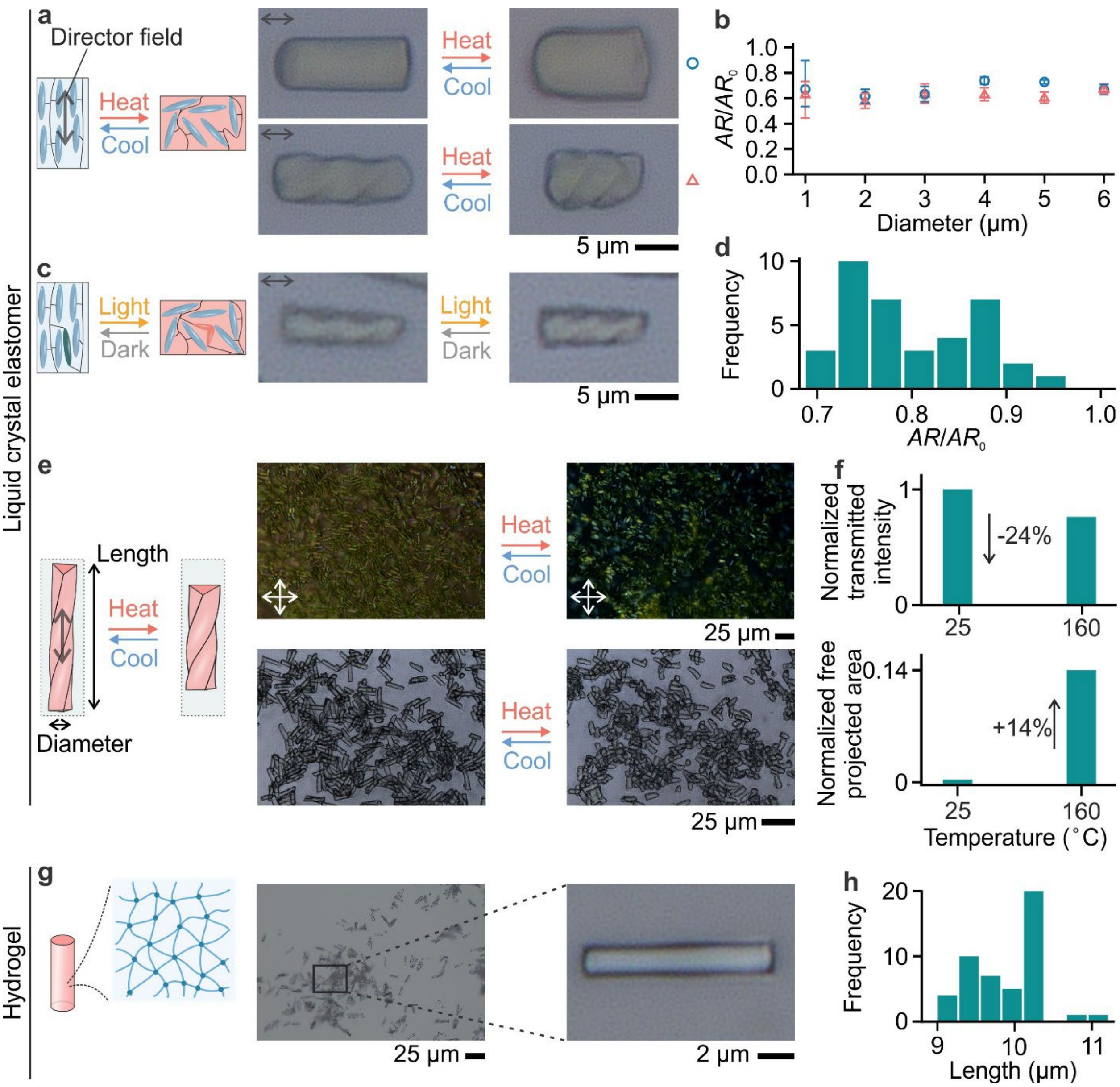

**Figure 4. Stimuli-responsive shape-shifting particles. a**, Optical microscopy images of cylindrical and chiral LCE particles before and after thermal actuation, showing reversible anisotropic deformation induced by heating. Heating drives an ordered nematic to a disordered isotropic transition within the elastomer network, resulting in reversible shape transformations. A double-headed arrow shows the director field, representing the average molecular alignment direction. **b**, Normalized aspect ratio ($AR/AR_0$) of cylindrical and chiral LCE particles with diameters ranging from 1–6 μm and fixed length of 15 μm during thermal actuation, where $AR_0$ denotes the initial particle aspect ratio prior to actuation. **c**, Ultraviolet-induced actuation in chiral LCE

particles, enabled by light-induced reversible trans–cis photoisomerization of the azobenzene crosslinkers, which locally disrupts nematic order within the LCE network. **d**, Normalized aspect ratio of chiral photoresponsive LCE particles (diameter 3 μm, length 15 μm) under ultraviolet illumination. **e**, Crossed-polarizer microscope images of chiral LCE particle suspensions before and after thermal actuation. White arrows indicate the polarization directions. **f**, Our results show an approximately 24% reduction in optical intensity under crossed polarizers in (**e**); and for a chiral particle with an initial length of 15 μm and a diameter of 3 μm, thermal actuation results in approximately 14% increase in open projected area within the field of view when lying parallel to the substrate. **g**, Optical microscope images of cylindrical hydrogel particles in a dehydrated state. **h**, The cylindrical hydrogel particles exhibited dimensions of 9.92 ± 0.46 μm in length, compared with the intended dimensions of 10 μm.

## 3. Discussion

We have established a general fabrication framework that independently controls particle geometry and material composition. By combining reusable three-dimensional master templates, soft lithography, swelling-assisted extraction, and sacrificial adhesive transfer, we fabricate colloidal particles with complex shapes and high geometric fidelity, with less than 8% deviation from the intended design. Beyond demonstrating a fabrication strategy, we establish a geometry-dependent extraction framework that relates particle shape and solvent-induced mold swelling, thereby defining the accessible design space of complex three-dimensional colloidal architectures.

A key feature of our approach is its ability to expand the accessible design space of colloidal particles, including longitudinal lengths, cross-sectional anisotropy, polygonal faceting, chirality, and longitudinal cross-section gradients. Beyond free-standing particles, our approach enables the fabrication of high-aspect-ratio pillar arrays with three-dimensional geometries. These surfaces may introduce additional control parameters through the geometry of the pillars in the field of biomimetic artificial active cilia[36-40], where reproducing the structure and kinematic complexity of natural cilia[37] still remains challenging[36,40]. These surfaces offer promising directions for fluid manipulation and transport control[36,39-41].

Our results offer a design framework governing the extraction process of the particles embedded within the PDMS mold, which is dictated by the interplay among particle geometry, particle-mold interfacial adhesion, and solvent-induced PDMS swelling. To distinguish between purely adhesion-mediated particle release and extraction involving deformation of the PDMS mold, we define the dimensionless parameter $d/h$ as the governing criterion, where $d$ is the geometric interlocking of the particle and $h$ is the swelling-induced release gap. When $d/h \leq 1$, particle extraction can proceed primarily through adhesive release, whereas when $d/h > 1$, extraction is expected to involve deformation of the surrounding PDMS mold. For the PDMS-hexane system studied here, we measured the criterion $d/h \leq 1$ to be $d < 0.32r$, where $r$ is the characteristic particle radius. More broadly, the framework enables assessment of particle extractability before fabrication, providing quantitative guidance for the design of increasingly complex particle geometries.

Additionally, our approach is compatible with different classes of stimuli-responsive materials. By fabricating particles from LCEs and hydrogels, we demonstrate that the strategy is not intrinsically restricted to a specific material system or actuation mechanism. Cylindrical and chiral LCE particles undergo reversible deformation under thermal stimuli, reducing their aspect ratio to approximately 60% of its original value when the nematic director is encoded parallel to the particle's long axis while preserving the overall three-dimensional architecture. Azobenzene-based photoresponsive units in the same LCE enable light-induced changes in particle geometry, resulting in a reduction of its aspect ratio to approximately 80% of its original value. Beyond the materials classes demonstrated here, the ability to independently program particle geometry and composition could be extended to include other functional materials, such as plasmonic nanoparticles[42-43], magnetic nanoparticles[44], catalytic materials[45], thereby enabling optical, magnetic, and chemical functionalities that are difficult to achieve using conventional colloidal fabrication approaches.

Our results also demonstrate that particle level responsive behavior can generate collective optical responses arising from coupled changes in the programmed liquid crystal organization within the particles and their geometry. Such coupling gives rise to distinct collective optical properties under different illumination conditions. Overall, this framework provides a route for independently engineering geometry and functionality in colloidal matter, establishing a novel direction for realizing programmable fluids, adaptive materials, and microscale robotic systems.

**Supporting Information**

Supporting Information is not included in this arXiv version.

**Acknowledgments**

This work was supported by the Department of Energy under Award No. DE-SC0005247 and by the National Science Foundation (NSF) through the Harvard University Materials Research Science and Engineering Center (MRSEC) under Award No. DMR-2011754. This work was funded by the SNSF (Swiss National Science Foundation) under the project number P500PN_217686. This work was performed in part at the Harvard University Center for Nanoscale Systems (CNS); a member of the National Nanotechnology Coordinated Infrastructure Network (NNCI), which is supported by the National Science Foundation under NSF award no. ECCS-2025158. We thank the staff of the CNS at Harvard University, particularly Adam Graham, Bok Yeop Ahn, and Mughees Khan, for technical assistance with Nanoscribe fabrication, soft lithography, and scanning electron microscopy. This work was performed in part at the Wyss Institute for Biologically Inspired Engineering at Harvard University. We thank Binh Diec and Adam Murrison for technical support and assistance with the UpNano system. Finally, we thank Friedrich Stricker for helpful discussions and Kieran McDaniel for help with experiments.

**Conflict of Interest**

The authors declare no conflict of interest.

## Methods and materials

**Fabrication of three-dimensional master templates.** Three-dimensional master templates were fabricated on glass substrates using commercial two-photon polymerization instruments (Nanoscribe and UpNano). After printing, the samples were developed according to the manufacturer's protocols to remove uncrosslinked photoresist and subsequently dried under ambient conditions. Note that to ensure the strong attachment of the printed structures to the glass substrate, a base layer with a thickness of 500 nm is printed.

**PDMS inverse mold fabrication.** Inverse molds were fabricated by casting polydimethylsiloxane (PDMS) over the printed master templates. Prior to PDMS casting, the master structures were functionalized via vapor-phase deposition (for 12 hours) of Triethoxy(1*H*,1*H*,2*H*,2*H*-perfluoro-1-octyl) silane to reduce adhesion between the printed templates and the cured PDMS during demolding. PDMS prepolymer (Sylgard 184, Dow Corning) and curing agent were mixed at a 10:1 weight ratio. The PDMS was cured by placing the sample in the oven at a temperature of 70 °C for 4 hours.

**Fabrication of liquid crystal elastomer structures.** The precursor formulation consisted of a side-on liquid crystal mesogen ((4''-acryloyloxybutyl) 2,5-di(4'-butyloxybenzoyloxy)benzoate), an azobenzene-based photoresponsive crosslinker (4,4'-bis(9-acryloyloxynonyloxy) azobenzene), and the photoinitiator IRGACURE® 819 (bis(2,4,6-trimethylbenzoyl)-phenylphosphine oxide), following the protocol in Ref. 34. The precursor formulation consisted of 90.5 mg mesogen, 7.5 mg crosslinker, and 2 mg photoinitiator, corresponding to 7.7 mol% crosslinker and 3.0 mol% photoinitiator. The precursor components were dissolved in dichloromethane and mixed using a vortex mixer to obtain a homogeneous liquid crystal precursor solution. The solvent was later removed using rotary evaporation. The prepared precursor mixture was deposited into the PDMS inverse molds and heated above the nematic-to-isotropic transition temperature to 110 °C to reduce viscosity and facilitate capillary infiltration of the liquid precursor into the embedded mold. The infiltration process was performed in a vacuum oven to ensure complete filling of PDMS mold cavities. A flat glass was subsequently placed on top of the PDMS mold to promote uniform filling and provide support for the LCE materials for the fabrication process (shearing). To encode internal liquid crystal order within the fabricated structures, the precursor-filled PDMS molds were placed within an externally applied magnetic field generated using neodymium permanent magnets (~0.25 T) during the alignment and polymerization process. The sample was subsequently cooled slowly down from 110 °C (1 °C $min^{-1}$) into the nematic phase at a temperature of 60 °C under the applied magnetic field and was kept at a temperature of 60 °C for an additional hour. After alignment, ultraviolet polymerization was carried out for 1 hour under an inert atmosphere to crosslink the reactive molecules and lock the aligned nematic director within the resulting LCE microstructures. For polymerization, we used a Dymax 2000-EC UV curing flood lamp with an illumination intensity of approximately 18 mW $cm^{-2}$. To minimize azobenzene trans–cis photoisomerization and preserve molecular alignment during curing, a 400 nm long-pass filter blocking wavelengths below 400 nm was used during polymerization. The resulting LCE is expected to exhibit a glass transition temperature of approximately 45 °C and a nematic-to-isotropic transition temperature of approximately 92 °C (see Ref. 34).

**Fabrication of hydrogel structures.** We prepared hydrogel precursor solutions using 2-hydroxyethyl methacrylate (HEMA) monomer (97.1 wt%), ethylene glycol dimethacrylate (EGDMA) crosslinker (1.9 wt%), and Darocur® 1173 (2-hydroxy-2-methylpropiophenone) photoinitiator (1.0 wt%). The precursor solution was then mixed using a vortex mixer and introduced into the PDMS inverse molds by capillary infiltration inside a vacuum chamber. After infiltration, the filled molds were covered with a glass substrate. The precursor-filled molds were subsequently exposed to ultraviolet illumination to initiate photopolymerization.

**Extraction and release of free-standing particles.** To release the fabricated particles from the PDMS molds, the samples were immersed in hexane and subjected to sonication to initiate swelling of the PDMS matrix for around 1 hour. Following swelling, a water-soluble adhesive tape (3M™) was pressed onto the swollen PDMS surface and then removed to transfer the particles from the PDMS mold onto the tape. The particles were subsequently released into aqueous solution by dissolving the tape in water.

**Swelling characterization of PDMS.** To characterize solvent-induced swelling of the PDMS molds, PDMS samples containing microstructured cavities were immersed in hexane for controlled time intervals. After swelling, the samples were immediately imaged in the solvent-swollen state using optical microscopy (Zeiss Microscopy). The characteristic swelling length was measured from the measured dimensional expansion of the PDMS mold.

**Thermal and optical actuation of liquid crystal elastomer particles.** Thermally induced actuation experiments were performed by heating the LCE particles, which are dispersed inside silicone oil (10,000 cSt), above the nematic-to-isotropic transition temperature at 160 °C while imaging the structures under bright-field and crossed-polarizer microscopy. Light-induced actuation experiments were performed using a 365 nm ultraviolet LED source (Thorlabs M365F1), which induced reversible trans–cis photoisomerization of the azobenzene crosslinkers. During optical actuation, the samples were kept at a temperature of 75 °C, above the glass transition temperature of LCE. The applied UV intensity was 115 mW $cm^{-2}$. Similarly, during optical actuation, the particles are dispersed in silicone oil with a viscosity of 10,000 cSt.

**Microscopy and particle characterization.** Scanning electron microscopy (Zeiss Gemini 360 FE-SEM) was used to characterize the geometry of fabricated structures prior to release. The free-standing particles images and quantitative measurements (all data except Figure 1c-e) were measured from microscopy images (KEYENCE VHX-X1, 500X–2500X objectives) using custom image-analysis code. The Data in Figure 1c-e are captured using a Zeiss Microscope. Note that mean-squared displacement (MSD) values were calculated from corrected particle trajectories to isolate Brownian motion. Background convective motion was removed by estimating a spatially varying linear flow field from particle displacements between consecutive frames and subtracting the predicted flow from the measured frame-to-frame particle displacements before reconstructing the corrected trajectories and calculating the MSD.

## References

1. Glotzer, S. C. & Solomon, M. J. Anisotropy of building blocks and their assembly into complex structures. *Nature Materials* **6**, 557-562 (2007).

2. Sacanna, S. & Pine, D. J. Shape-anisotropic colloids: Building blocks for complex assemblies. *Current Opinion in Colloid & Interface Science* **16**, 96-105 (2011).

3. Almohammadi, H., Khadem, S. A., Azzari, P., Yuan, Y., Guerra, A., Rey, A.D. & Mezzenga, R. Liquid–liquid crystalline phase separation of filamentous colloids and semiflexible polymers: experiments, theory and simulations. *Reports on Progress in Physics* **88**, 036601 (2025).

4. Tanjeem, N., Minnis, M. B., Hayward, R. C. & Shields IV, C. W. Shape-changing particles: from materials design and mechanisms to implementation. *Advanced Materials* **34**, 2105758 (2022).

5. Djellouli, A., Van Raemdonck, B., Wang, Y., Yang, Y., Caillaud, A., Weitz, D., Rubinstein, S., Gorissen, B. & Bertoldi, K. Shell buckling for programmable metafluids. *Nature* **628**, 545–550 (2024).

6. Chen, C., Martínez Narváez, C.D., Chang, N., Jimenez, C.M., Dennis, J.M., Jaeger, H.M., Rowan, S.J. & de Pablo, J.J. Tunable shear thickening, aging, and rejuvenation in suspensions of shape-memory-endowed liquid crystalline particles. *Proceedings of the National Academy of Sciences* **122**, e2425373122 (2025).

7. Damasceno, P. F., Engel, M. & Glotzer, S. C. Predictive self-assembly of polyhedra into complex structures. *Science* **337**, 453-457 (2012).

8. Zeravcic, Z., Manoharan, V. N. & Brenner, M. P. Size limits of self-assembled colloidal structures made using specific interactions. *Proceedings of the National Academy of Sciences* **111**, 15918-15923 (2014).

9. Abdelrahman, M. K., Wagner, R. J., Kalairaj, M. S., Zadan, M., Kim, M. H., Jang, L. K.,Wang, S., Javed, M., Dana, A., Singh, K. A., Hargett, S. E., Gaharwarm A. K., Majidi, C., Vernerey, F. J. & Ware, T. H. Material assembly from collective action of shape-changing polymers. *Nature Materials* **23**, 281–289 (2024).

10. Nguyen, T. D., Jankowski, E. & Glotzer, S. C. Self-assembly and reconfigurability of shape-shifting particles. *ACS Nano* **5**, 8892–8903 (2011).

11. Gang, O. & Zhang, Y. Shaping phases by phasing shapes. *ACS Nano* **5**, 8459–8465 (2011).

12. Yoo, J. W. & Mitragotri, S. Polymer particles that switch shape in response to a stimulus. *Proceedings of the National Academy of Sciences* **107**, 11205-11210 (2010).

13. Champion, J. A., Katare, Y. K. & Mitragotri, S. Making polymeric micro-and nanoparticles of complex shapes. *Proceedings of the National Academy of Sciences* **104**, 11901-11904 (2007).

14. Gong, T., Zhao, K., Wang, W., Chen, H., Wang, L. & Zhou, S. Thermally activated reversible shape switch of polymer particles. *Journal of Materials Chemistry B* **2**, 6855-6866 (2014).

15. Wischke, C., Schossig, M. & Lendlein, A. Shape-memory effect of micro-/nanoparticles from thermoplastic multiblock copolymers. *Small* **10**, 83-87 (2013).

16. Lee, K. J., Yoon, J., Rahmani, S., Hwang, S., Bhaskar, S., Mitragotri, S. & Lahann, J. Spontaneous shape reconfigurations in multicompartmental microcylinders. *Proceedings of the National Academy of Sciences* **109**, 16057-16062 (2012).

17. Sitt, A., Soukupova, J., Miller, D., Verdi, D., Zboril, R., Hess, H. & Lahann, J. Microscale rockets and picoliter containers engineered from electrospun polymeric microtubes. *Small* **12**, 1432-1439 (2016).

18. Ohm, C., Kapernaum, N., Nonnenmacher, D., Giesselmann, F., Serra, C. & Zentel, R. Microfluidic synthesis of highly shape-anisotropic particles from liquid crystalline elastomers with defined director field configurations. *Journal of the American Chemical Society* **133**, 5305-5311 (2011).

19. Hessberger, T., Braun, L. & Zentel, R. Microfluidic synthesis of actuating microparticles from a thiol-ene based main-chain liquid crystalline elastomer. *Polymers* **8**, 410 (2016).

20. Ohm, C., Serra, C. & Zentel, R. A Continuous flow synthesis of micrometer-sized actuators from liquid crystalline elastomers. *Advanced Materials* **21**, 4859-4862 (2009).

21. Hessberger, T., Braun, L. B., Henrich, F., Müller, C., Gießelmann, F., Serra, C. & Zentel, R. Co-flow microfluidic synthesis of liquid crystalline actuating Janus particles. *Journal of Materials Chemistry C* **4**, 8778-8786 (2016).

22. Wilborn, M., Almohammadi, H., Qu, P., Wang, Y., Kay, R., Kim, D., Bertoldi, K., Weitz, D. & Aizenberg, J. Towards differentiation in untethered microactuators: a soft fabrication strategy. *Advanced Materials* **37**, 2507273 (2025).

23. Buguin, A., Li, M. H., Silberzan, P., Ladoux, B. & Keller, P. Micro-actuators: When artificial muscles made of nematic liquid crystal elastomers meet soft lithography. *Journal of the American Chemical Society* **128**, 1088-1089 (2006).

24. Epstein, E., Yoon, J., Madhukar, A., Hsia, K. J.& Braun, P. V. Colloidal particles that rapidly change shape via elastic instabilities. *Small* **11**, 6051–6057 (2015).

25. Kawata, S., Sun, H.-B., Tanaka, T. & Takada, K. Finer features for functional microdevices. *Nature* **412**, 697–698 (2001).

26. Guo, Y., Shahsavan, H. & Sitti, M. 3D microstructures of liquid crystal networks with programmed voxelated director fields. *Advanced Materials* **32**, 2002753 (2020).

27. Maruo, S., Nakamura, O. & Kawata, S. Three-dimensional microfabrication with two-photon-absorbed photopolymerization. *Optics Letters* **22**, 132-134 (1997).

28. Liu, S. F., Hou, Z. W., Lin, L., Li, Z. & Sun, H. B. 3D laser nanoprinting of functional materials. *Advanced Functional Materials* **33**, 2211280 (2023).

29. White, T. J. & Broer, D. J. Programmable and adaptive mechanics with liquid crystal polymer networks and elastomers. *Nature Materials* **14**, 1087– 1098 (2015).

30. Herbert, K. M., Fowler, H. E., McCracken, J. M., Schlafmann, K. R., Koch, J. A. & White, T. J. Synthesis and alignment of liquid crystalline elastomers. *Nature Reviews Materials* **7**, 23–38 (2021).

31. Yao, Y., Wilborn, A. M., Lemaire, B., Trigka, F., Stricker, F., Weible, A. H., Li, S., Bennett, R. K., Cheung, T. C., Grinthal, A., Zhernenkov, M., Freychet, G., Wąsik, P., Kozinsky, B., Lerch, M. M., Wang, X. & Aizenberg, J. Programming liquid crystal elastomers for multistep ambidirectional deformability. *Science* **386**, 1161–1168 (2024).

32. Lee, J. N., Park, C. & Whitesides, G. M. Solvent compatibility of poly (dimethylsiloxane)-based microfluidic devices. *Analytical chemistry* **75**, 6544-6554 (2003).

33. Yao, Y., Waters, J.T., Shneidman, A.V., Cui, J., Wang, X., Mandsberg, N.K., Li, S., Balazs, A.C. & Aizenberg, J. Multiresponsive polymeric microstructures with encoded predetermined and self-regulated deformability. *Proceedings of the National Academy of Sciences* **115**, 12950-12955 (2018).

34. Li, S., Aizenberg, M., Lerch, M. M. & Aizenberg, J. Programming deformations of 3D microstructures: Opportunities enabled by magnetic alignment of liquid crystalline elastomers. *Accounts of Materials Research* **4**, 1008–1019 (2023).

35. Peppas, N. A., Moynihan, H. J. & Lucht, L. M. The structure of highly crosslinked poly (2-hydroxyethyl methacrylate) hydrogels. *Journal of biomedical materials research* **19**, 397-411 (1985).

36. Mathijssen, A., Almohammadi, H., Altman, L., Calazans, T., Ferencz, M. J., Fung, M., Lee, I. J., Lisicki, M., Liu, I., Liu, M., Liu, T., Park, E., Tao, R., Thery, A., Wang, Z. & Young, M. Biomedical active matter: Emergence and breakdown of collective functionalities. *arXiv:2603.15778* (2026).

37. Ramirez-San Juan, G. R., Mathijssen, A. J., He, M., Jan, L., Marshall, W. & Prakash, M. Multi-scale spatial heterogeneity enhances particle clearance in airway ciliary arrays. *Nature Physics* **16**, 958-964 (2020).

38. Li, S., Lerch, M. M., Waters, J. T., Deng, B., Martens, R. S., Yao, Y., Kim, D. Y., Bertoldi, K., Grinthal, A., Balazs, A. C. & Aizenberg, J. Self-regulated non-reciprocal motions in single-material microstructures. *Nature* **605**, 76–83 (2022).

39. Wang, W., Liu, Q., Tanasijevic, I., Reynolds, M.F., Cortese, A.J., Miskin, M.Z., Cao, M.C., Muller, D.A., Molnar, A.C., Lauga, E. & McEuen, P.L. Cilia metasurfaces for electronically programmable microfluidic manipulation. *Nature* **605**, 681-686 (2022).

40. Peerlinck, S., Milana, E., De Smet, E., De Volder, M., Reynaerts, D. & Gorissen, B. Artificial cilia–bridging the gap with nature. *Advanced Functional Materials* **33**, 2300856 (2023).

41. Park, J., Moon, C.S., Lee, J.M., Rahat, S.A., Kim, S.M., Pham, J.T., Kappl, M., Butt, H.J. & Wooh, S. Bioinspired capillary force-driven super-adhesive filter. *Nature* **643**, 388–394 (2025).

42. Sun, Y., Evans, J. S., Lee, T., Senyuk, B., Keller, P., He, S. & Smalyukh, I. I. Optical manipulation of shape-morphing elastomeric liquid crystal microparticles doped with gold nanocrystals. *Applied Physics Letters* **100**, 241901 (2012).

43. Liu, X., Wei, R., Hoang, P. T., Wang, X., Liu, T. & Keller, P. Reversible and rapid laser actuation of liquid crystalline elastomer micropillars with inclusion of gold nanoparticles. *Advanced Functional Materials* **25**, 3022-3032 (2015).

44. Erb, R. M., Son, H. S., Samanta, B., Rotello, V. M. & Yellen, B. B. Magnetic assembly of colloidal superstructures with multipole symmetry. *Nature* **457**, 999-1002 (2009).

45. Paxton, W. F., Kistler, K. C., Olmeda, C. C., Sen, A., St. Angelo, S. K., Cao, Y., Mallouk, T.E., Lammert, P.E. & Crespi, V.H. Catalytic nanomotors: autonomous movement of striped nanorods. *Journal of the American Chemical Society* **126**, 13424-13431 (2004).